\documentclass[twocolumn,aps,floatfix,pra]{revtex4}
\usepackage{amsmath}
\usepackage{graphicx}
 
\begin{document}

\title{Dynamics of a few interacting bosons escaping from an open well}
\author{Jacek Dobrzyniecki}
\author{Tomasz Sowi\'nski}
\affiliation{                    
   Institute of Physics, Polish Academy of Sciences, Aleja Lotnikow 32/46, PL-02668 Warsaw, Poland
  }

\begin{abstract}
The dynamics of a few ultra-cold bosons tunneling from a one-dimensional potential well into an open space is studied. In such a system several decay channels can be distinguished, each corresponding to a different number of bosons escaping simultaneously. We show that as the inter-particle interaction strength is changed, the system undergoes transitions between distinct regimes characterized by the dominance of different decay channels. These transitions are reflected in the behavior of the decay rate of the system, which is measurable experimentally. By means of a simple theoretical description, we show that the transitions occur at the points where a new decay channel becomes energetically viable. The results provide insight into the behavior of decaying few-body systems and may have potential interest for experiments. 
\end{abstract}

\maketitle
\section{Introduction}
Tunneling through a classically impenetrable barrier is a hallmark effect of quantum mechanics. In Gamow's seminal work from 1928 \cite{gamow1928}, a quantum tunneling was used to explain the phenomenon of $\alpha$-decay that resisted a satisfactory classical explanation. Gamow expressed the problem in terms of a particle escaping from a finite potential well into open space. The model of particles tunneling out of a potential trap has since seen wide use in the analysis of various phenomena in physics, including proton emission \cite{talou1999,talou2000}, fusion, fission, photoassociation and photodissociation \cite{keller1984,bhandari1991,balatenkin1998,vatasescu2000}. While the escape behavior of a single particle is well understood theoretically \cite{razavy2003}, and the tunneling of a dilute Bose-Einstein condensate of a large number of particles is appropriately captured in the mean-field approximation \cite{salasnich2001,carr2005,huhtamaki2007,zhao2017}, a thorough description of the dynamics of interacting few-body systems remains elusive \cite{lode2012,hunn2013}. While extensive work has been done on the dynamics of bosonic systems tunneling between individual sites of an optical lattice (see e.g. \cite{keshavamurthy,neuhaussteinmetz2017} and the citations therein), the dynamics in open systems have received comparatively less attention.

Thanks to brand new developments in the field of ultra-cold physics, the few-body tunneling problem has seen significant interest in recent years  \cite{delcampo2006,lode2009,kim2011,taniguchi2011,lode2012,maruyama2012,hunn2013,lode2014,maksimov2014,gharashi2015,lundmark2015,lode2015,ishmukhamedov2017}. New techniques give the experimentalist precise control over the potential landscape \cite{meyrath2005,henderson2009,vanes2010}, effective dimensionality \cite{gorlitz2001,greiner2001,schreck2001}, initial state \cite{serwane2011} and inter-particle interactions \cite{chin2010,pethick2008,zollner2008a,zollner2008b}. Notable experiments in the area have been done by the Heidelberg group, where the decay of a system of a few distinguishable fermions was studied \cite{zurn2012,zurn2013}. 

It is known that the decay of trapped few-body systems can take place via several different processes. For example, in the case of a trapped two-body system, the particles may tunnel sequentially, one by one, or they may escape simultaneously as a bound pair \cite{rontani2012,rontani2013}. An interesting question is the relative contribution of the different tunneling channels to the overall decay process. A few works have touched on the question with regards to two-body systems \cite{maruyama2012,gharashi2015,lundmark2015,zurn2013}. However, a systematic treatment, specifically when systems with more than two particles are considered, is still missing. 

In this work, we aim to qualitatively analyze the few-body decay processes and investigate the contribution of separate decay channels. We numerically simulate a one-dimensional system of a few (two and three) bosons, escaping from a potential well into open space. We investigate how the nature of the decay changes with the interaction strength. Our focus is on attractive forces, which, due to energy conditions, strongly support many-body tunneling and suppress sequential tunneling. A simple model is provided for estimating what tunneling mechanisms are available in different interaction regimes. We show that predictions of this oversimplified model surprisingly well reflect the results of the numerical simulations. Our work is complementary to the results presented in \cite{zurn2013,rontani2013,gharashi2015} where distinguishable fermions were studied.

The work is organized as follows. In Section \ref{sec:model} we describe the model system under study. In Section \ref{sec:2boson} we describe the decay dynamics of a two-boson system, while also establishing a toolbox of techniques that allow to closely analyze the structure of the decay process. In Section \ref{sec:2bosonlong} we focus on the long-time dynamics of the two-boson system. In Section \ref{sec:3boson} we describe the decay of a three-boson system. In Section \ref{sec:treatment} we give the results a theoretical foundation by describing a simple model of few-body decay and showing that its predictions agree well with the numerical results. In Section \ref{sec:potential-shape} we analyze the influence of the shape of the potential beyond the barrier on the dynamics. Section \ref{sec:conclusion} is the conclusion.

\section{The model}
\label{sec:model}

We consider an ultra-cold system of $N$ indistinguishable bosons of mass $m$, confined in a one-dimensional external trap. The many-body Hamiltonian of the system reads
\begin{equation}
\label{eq:mb_hamiltonian_1}
    H = \sum_i \left[ -\frac{\hbar^2}{2m} \frac{\partial^2}{\partial{x_i}^2} + V(x_i) \right] + g\sum_{i<j} \delta(x_i-x_j),
\end{equation}
where $x_i$ represents the position of the $i$-th boson. The interaction potential is modeled by the $\delta$ function. This is a good approximation for ultra-cold particles, for which the de Broglie wavelength is larger than the average inter-particle distance and fine spatial details of scattering beyond the $s$-wave level can be ignored. The inter-particle interaction strength $g$ is related to the $s$-wave scattering length \cite{olshanii1998,haller2009} and in experiments it can be tuned via the Feshbach resonance technique \cite{chin2010,pethick2008} or by changes of the shape of the confinement in perpendicular directions \cite{olshanii1998}. 

\begin{figure}
\includegraphics[width=1\linewidth]{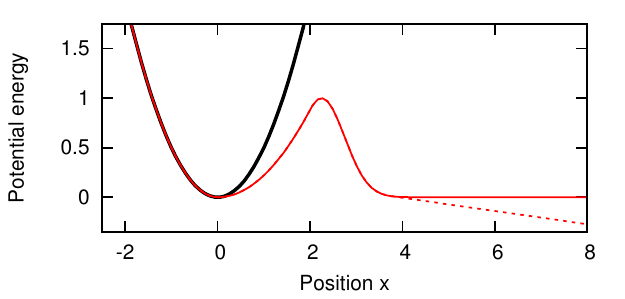}
\caption{The trap potential $V(x)$ as defined by the equation (\ref{eq:potential}) (red solid line) and the modified potential $V'(x)$ defined by the equation (\ref{eq:potential-unbounded}) (red dotted line). Note that $V(x)$ and $V'(x)$ are identical for $x < 3.75$. The initial state of the system is prepared as the interacting ground state in a harmonic oscillator $\frac{1}{2} m \Omega_0^2 x^2$ (thick line). Energy and length are given in units of $\hbar\Omega_0$ and $\sqrt{\hbar/m\Omega_0}$, respectively.}
 \label{Fig1} 
\end{figure}

For convenience we assume that initially particles are confined in a harmonic trap of frequency $\Omega_0$. Then the trap is suddenly opened and it has the following form ($x_0 = \sqrt{\hbar/m\Omega_0}$ is the harmonic oscillator length unit):
\begin{equation}
\label{eq:potential}
 V(x) = 
 \begin{cases}
    m\Omega_0 x^2/2 ,& x< 0 \\   
    m\Omega x^2/2 ,& 0 \le x\le 2x_0\\   
  \hbar\Omega_0 \mathrm{e}^{-2(x/x_0-9/4)^2} ,& x > 2x_0
\end{cases}
\end{equation}
which corresponds to a potential well, separated from free space by a finite potential barrier. The modified frequency $\Omega\approx\Omega_0/2.26$ ensures that the potential shape $V(x)$ is a continuous function and has a continuous first derivative in the entire space (see Fig.~\ref{Fig1}). To make the analysis more complete, later (in Section \ref{sec:potential-shape}) we also study the influence of the shape of the potential beyond the barrier. Therefore we examine a small modification of the external potential with linear ramping outside the well:
\begin{equation}
\label{eq:potential-unbounded}
 V'(x) = 
 \begin{cases}
    V(x) ,& x \le (15/4)x_0 \\   
  \hbar\Omega_0 \left[6(x/x_0-15/4)+1\right]/90 ,& x > (15/4)x_0.
\end{cases}
\end{equation}
Coefficients are chosen so that the function $V'(x)$ and its first derivative are continuous everywhere. This modified potential is shown in Fig.~\ref{Fig1} as a dotted line. 

In the following, we express all quantities in natural harmonic oscillator units, {\it i.e.} energy, length, and the interaction are given in $\hbar \Omega_0$, $\sqrt{\hbar/m\Omega_0}$, and $\sqrt{\hbar^3\Omega_0/m}$, respectively. 

Since initially all particles are confined in the harmonic trap, the initial many-body state of the system $|\Psi_0\rangle$ is identical to the ground state of $N$ interacting bosons in this trap. For vanishing interactions ($g=0$) it is a simple product state $|\Psi_0\rangle=\left(\hat{a}_0^\dagger\right)^N|\mathtt{vac}\rangle$, where $\hat{a}_0$ annihilates a boson in the ground-state orbital of the corresponding single-particle problem. For non-vanishing interactions ($g\ne0$) the $N$-body ground state $|\Psi_0\rangle$ is found numerically by propagating a trial wave function in the imaginary time. The evolution of the system for $t>0$ is performed straightforwardly by solving a time-dependent many-body Schr\"odinger equation in position representation. These calculations are done on a dense discrete grid taking into account a huge extent of space in the region where the potential vanishes. For the $N=2$ case $x\in [-10; 90]$ (in natural units of a harmonic oscillator), while for the $N = 3$ case $x \in [-4; 40]$. We have verified that the chosen numerical parameters are sufficient for the short-time scales considered, {\it i.e.}, enlarging the system does not significantly affect the results obtained.

\section{Two bosons case}
\label{sec:2boson}

First, we focus on the simplest nontrivial system of $N = 2$ interacting bosons. After the sudden change of the potential at $t=0$ the particles start to escape to the open space through the barrier. The resulting dynamical properties of the many-body interacting system can be well described in the language of appropriate probabilities. In the case of two bosons one distinguishes three different probabilities which can be quite easily measured experimentally \cite{zurn2013,rontani2013}: the probability that both particles remain in the trap ${\cal P}_2(t)$, the probability that exactly one particle occupies the trap ${\cal P}_1(t)$, and the probability that exactly both particles are out of the trap ${\cal P}_0(t)$. These probabilities are directly encoded in the two-particle density profile $\rho(x_1,x_2,t)=|\Psi(x_1,x_2,t)|^2$ as appropriate integrals
\begin{equation}
  {\cal P}_k(t) = \int_{\mathbf{P}_i}\!\!\mathrm{d}x_1 \mathrm{d}x_2\,\,\rho(x_1,x_2,t).
\end{equation}
Integrations are performed over appropriate regions $\mathbf{P}_i$ of two-particle positions, {\it i.e.}, $x_1,x_2<x_w$ for $\mathbf{P}_2$, $x_1,x_2>x_w$ for $\mathbf{P}_0$, and remaining configurations for $\mathbf{P}_1$, where $x_w\approx 3x_0$ is the position of the barrier. Of course initially one finds ${\cal P}_2(0)=1$ and ${\cal P}_1(0)={\cal P}_0(0)=0$. 

 \begin{figure}
 \includegraphics[width=1\linewidth]{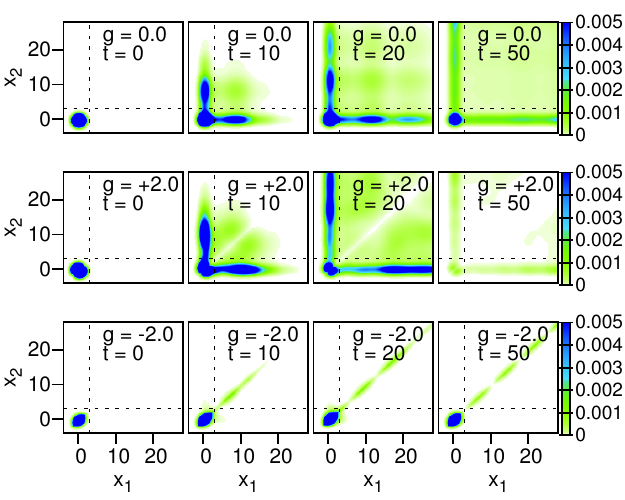}
 \caption{Time evolution of the density distribution $\rho(x_1,x_2,t)$ in the two-boson system with different interaction strengths $g$. The dashed lines demarcate the well boundary $x_w\approx3x_0$. For the non-interacting and repulsive systems ($g = 0, g = 2$) only sequential tunneling is present, while in the strongly attractive system ($g = -2$) essentially the entire decay process takes place via pair tunneling. Positions are in units of $\sqrt{\hbar/m\Omega_0}$, interaction strength in units of $\sqrt{\hbar^3\Omega_0/m}$, time in units of $1/\Omega_0$.}
 \label{Fig2} 
\end{figure}

In Fig.~\ref{Fig2} we show a time evolution of the two-particle density profile $\rho(x_1,x_2,t)$ for different interactions between particles. The dashed lines, located at $x_w$, divide the configuration space into three different regions $\mathbf{P}_i$. At the beginning ($t = 0$) the many-body wave function is located only within the region $\mathbf{P}_2$. For a trapped system of two bosons, there are essentially two mechanisms of particle loss from the trap: bosons can tunnel out sequentially (one after the other), or they can tunnel simultaneously as a bound pair. 

In the non-interacting case ($g=0$, top row in Fig.~\ref{Fig2}) the two particles tunnel completely independently. A significant amount of probability density flows from $\mathbf{P}_2$ into the $\mathbf{P}_1$ region. This is a signature of sequential tunneling, where one boson has tunneled out while the other still remains in the well. Note the visible oscillations in the probability density flowing into $\mathbf{P}_1$. They appear because, after the potential landscape is changed at $t = 0$, the initial wave function is no longer the ground state of the Hamiltonian and the density begins oscillating back and forth inside the well.

There is also a non-negligible concentration of probability density within the $\mathbf{P}_0$ region, corresponding to both bosons having tunneled out. Due to the absence of interactions, the two-particle density shows no correlations, {\it i.e.}, the two-particle wave function is a simple product of two identical one-particle wave functions. 

For repulsive interactions ($g=2$, middle row) the sequential tunneling is enhanced. The probability first flows from $\mathbf{P}_2$ into the $\mathbf{P}_1$ region, and subsequently begins to flow from the areas of increased density in $\mathbf{P}_1$ into $\mathbf{P}_0$, corresponding to the second boson escaping the well (when the first one is already outside). Moreover, we observe a vanishing two-particle density along the $x_1=x_2$ diagonal in the $\mathbf{P}_0$ region. It means that the pair tunneling (correlated in positions) is almost completely suppressed.

The situation is markedly different for a strongly attractive system ($g = -2$, bottom row). As it is seen, in this case a pair tunneling is the only decay mechanism present, and the sequential tunneling is absent. Hence, the probability density within the $\mathbf{P}_1$ region practically vanishes. Instead, density flows from $\mathbf{P}_2$ directly into the $\mathbf{P}_0$ region and remains concentrated in $\mathbf{P}_0$ along the $x_1 = x_2$ diagonal, representing the bosons traveling together as a bound pair. 
 
 \begin{figure}
 \includegraphics[width=1\linewidth]{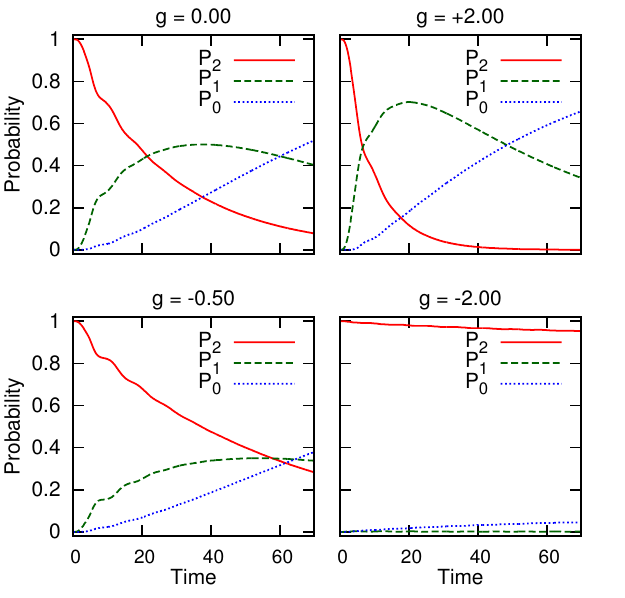}
 \caption{Time evolution of the probabilities ${\cal P}_k(t)$ of finding exactly $k$ particles inside the well, for the two-boson system with different interaction strengths $g$. For the cases $g = 0, g = 2$ and $g= -0.5$ the evolution is governed mainly by the two-stage process of sequential tunneling, in which the probability flows along ${\cal P}_2(t) \rightarrow {\cal P}_1(t) \rightarrow {\cal P}_0(t)$. For the strongly attractive system ($g = -2$) both bosons tunnel simultaneously, so that probability flows straight from ${\cal P}_2(t)$ to ${\cal P}_0(t)$. Time and interaction strength $g$ are given in units of $1/\Omega_0$ and $\sqrt{\hbar^3\Omega_0/m}$, respectively.}
 \label{Fig3}
\end{figure}

For a more quantitative description of the many-body tunneling process, in Fig.~\ref{Fig3} we show the time-dependence of the probabilities ${\cal P}_k(t)$ for different interactions.

For the non-interacting system ($g = 0$), the sequential decay of the system constitutes the dominant decay process. Hence the time evolution of the ${\cal P}_k(t)$ resembles a two-stage nuclear decay. ${\cal P}_2(t)$ steadily decreases until it reaches zero, ${\cal P}_1(t)$ grows at first but then reaches a maximum and begins decreasing, and ${\cal P}_0(t)$ steadily increases throughout the entire process. Since the potential supports no bound states, for $t \rightarrow \infty$ one expects ${\cal P}_2(t) = {\cal P}_1(t) = 0$ and ${\cal P}_0(t) = 1$. 

For a repulsive system ($g = 2.0$), as well as for a weakly attractive system ($g = -0.5$), the evolution of probabilities remains similar to the non-interacting case. In the repulsive case, the interaction causes the first tunneling boson to see a lower effective potential barrier. Accordingly, the depletion of ${\cal P}_2(t)$ is faster than in the non-interacting system. Furthermore, the single boson left in the trap feels no interaction and tunnels more slowly than the first. In addition, the probability of finding exactly one boson in the trap becomes higher, and ${\cal P}_1(t)$ reaches a higher maximum value. In the attractive case, the depletion of ${\cal P}_2(t)$ is slower and ${\cal P}_1(t)$ is smaller, for analogous reasons.

The dynamics becomes significantly different in the strongly attractive regime ($g = -2.0$). Here the role of the two-stage sequential decay is negligible. Therefore ${\cal P}_1(t)$ remains near zero at all times, while the depletion of ${\cal P}_2(t)$ is nearly mirrored by a corresponding increase in ${\cal P}_0(t)$. However, in this limit the decay of ${\cal P}_2(t)$ is very slow compared to the non-interacting system.

While the short-time dynamics already give a quite good view of the nature of the decay process, a more in-depth understanding of the dynamics requires simulating the evolution over longer timescales. This is especially relevant for strongly attractive systems, where the decay is very slow and the characteristic qualities of the dynamics only become visible over a long time. Simulating the system dynamics for longer times is however problematic, as the particles leaving the well eventually reach the boundary of the box, leading to nonphysical reflections that distort the results. To avoid this problem, we incorporate the simplest method used widely in the framework of numerical simulation for open systems. Namely, far from the trap we add an imaginary potential term $-i\Gamma(x)$ to the single-particle Hamiltonian, to absorb particles \cite{muga2004}. The form of $\Gamma(x)$ should be chosen carefully to minimize reflections and to ensure complete absorption far from the trap \cite{riss1996}. For this reason, we choose a smoothly rising function $\Gamma(x) = 10^{-3}\cdot(x-20)^2$ (for $x>20$ in natural units of harmonic oscillator). We checked that the final results do not depend on details of $\Gamma(x)$. 

Of course, with this simplified approach, one cannot predict all possible properties of the system. In particular, all quantities which depend on the microscopic state of the system in the region where the absorbing potential is present are strongly affected by this non-physical mechanism \cite{selsto2010}. However, quantities which depend only on the state of particles in the region far from the absorbing border, such as the probability ${\cal P}_2(t)$, are well captured by the evolution.

\section{Long-time dynamics of the two-boson system}
\label{sec:2bosonlong}

\begin{figure}
 \includegraphics[width=1\linewidth]{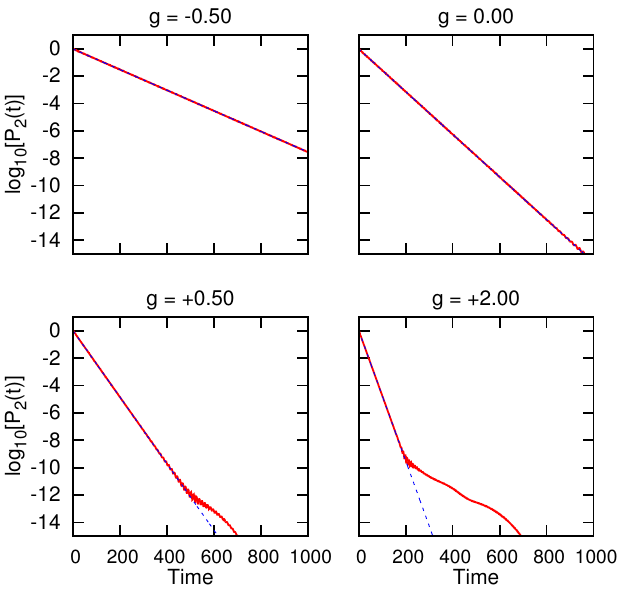}
 \caption{Time evolution of the probability ${\cal P}_2(t)$ over a long time scale (red, solid) for the two-boson system with various interaction strengths $g$. Blue dashed line shows an exponential fit to ${\cal P}_2(t)$. It can be seen that ${\cal P}_2(t)$ decays exponentially (apart from very long times). Time and interaction strength $g$ are given in units of $1/\Omega_0$ and $\sqrt{\hbar^3\Omega_0/m}$, respectively.}
 \label{Fig4} 
\end{figure}

It is known that for decaying systems such as the one under study, the survival probability (the probability ${\cal P}_S(t) = |\langle \Psi_0 | \Psi(t) \rangle|^2$ that the system remains in the initial state) obeys an exponential decay law to a very good approximation \cite{davydov1976}. For the two-body trapped system, the survival probability is mimicked by the probability ${\cal P}_2(t)$. Therefore, its time evolution should be approximately given by

\begin{equation}
    {\cal P}_2(t) \sim e^{-\gamma t}, 
\end{equation}
with some constant decay rate $\gamma$. To confirm this assumption, in Fig.~\ref{Fig4} we show the time evolution of the probability ${\cal P}_2(t)$ over a long time scale, along with an exponential fit. It can be seen that throughout the decay process the time evolution of ${\cal P}_2(t)$ indeed follows an exponential form. For very short and long times deviations from exponential decay are present. However, such nonexponential features are expected, since in a system with an energy spectrum bounded from below the decay cannot be exponential for all times \cite{khalfin1958}. Physically, the nonexponential decay can be interpreted as representing the possibility of reconstruction of the initial state from the decay fragments \cite{fonda1972,muga2006}. For the studied system, the short-time deviations are almost invisible on the scale of the plot. On the other hand, the long-time deviations appear when the trapped system is practically completely depleted and ${\cal P}_2(t)$ is quite negligible. Thus essentially the entire tunneling process is governed by an exponential regime. Accordingly, we can describe the decay through a single constant decay rate $\gamma$. 

Note that for sufficiently large times, the dynamics may be affected by small reflections from the complex absorbing potential. However, such fine details are unimportant for the overall problem studied in the work, since we focus only on the dynamics in the intermediate times where exponential decay dominates the dynamics. 

 \begin{figure}
 \includegraphics[width=1\linewidth]{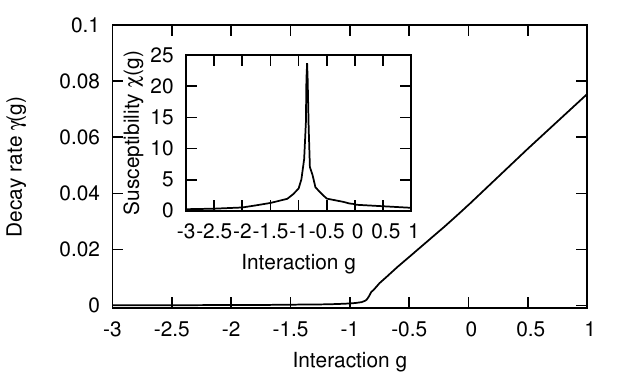}
 \caption{The decay rate $\gamma(g)$ of ${\cal P}_2(t)$ as a function of $g$, for the two-boson system. Inset: The susceptibility $\chi(g) = \gamma^{-1} (\partial \gamma / \partial g)$. A sudden change of the behavior is clearly seen at $g = -0.85$. Interaction strength $g$ is given in units of $\sqrt{\hbar^3\Omega_0/m}$, decay rate in units of $\Omega_0$, susceptibility in units of $\sqrt{m/\hbar^3\Omega_0}$.}
 \label{Fig5} 
\end{figure}

As shown in Fig.~\ref{Fig5}, the decay rate $\gamma$ essentially depends on interaction $g$. For convenience, we also show its susceptibility, defined as $\chi(g) = \gamma^{-1} (\partial \gamma / \partial g)$. The decay rate grows monotonically with $g$. However, a significant change occurs at $g \approx -0.85$. The growth of $\gamma(g)$ becomes significantly faster above this point, which is accompanied by a sharp, clear peak in $\chi(g)$. This behavior is in agreement with our previous results. As was discussed, in the strong attraction limit the decay of the system takes place solely by the pair tunneling. The sequential tunneling appears as we move closer to $g=0$. This observation suggests that the qualitative change in the dependence of $\gamma(g)$ on $g$ corresponds to the activation of the sequential tunneling channel. To verify this hypothesis, one needs a method of determining the relative participation of the different tunneling channels. One of the possibilities is to consider the probability fluxes instead of probabilities. 

By the continuity equation, the derivative $J(t) = -\partial {\cal P}_2(t) / \partial t $ is equal to the total flow of probability out of the $\mathbf{P}_2$ region. The total flow $J(t)$ can be decomposed into two independent parts that correspond to the two different channels of ${\cal P}_2(t)$ decay: 

\begin{equation}
    J(t) = J_1(t) + J_0(t), 
\end{equation}
where $J_i(t)$ is the probability flow between the $\mathbf{P}_2$ and $\mathbf{P}_i$ regions. For purpose of numerical calculations, $J(t)$ can be also calculated as a line integral

\begin{equation}
J(t) = \oint_{\partial\mathbf{P}_2} j(x_1,x_2;t) \mathrm{d}\mathbf{l},
\end{equation}
where $\mathrm{d}\mathbf{l}$ is a line element of the boundary $\partial\mathbf{P}_2$, and $j(x_1,x_2;t)$ is the outgoing probability flux through $\partial\mathbf{P}_2$. In this approach one divides the boundary $\partial\mathbf{P}_2$ into two segments $B_1$ and $B_0$, indicating a border between the region $\mathbf{P}_2$ and the regions $\mathbf{P}_1$ and $\mathbf{P}_0$, respectively. This allows us to calculate $J_1(t)$ and $J_0(t)$ separately, by integrating the outgoing probability flux only along the corresponding boundary segment. 

It should be noted that in this approach there is a tendency for slight overestimation of the flux $J_0(t)$. This is because in practice a rapid sequence of single-boson tunnelings cannot be distinguished from pair tunneling. However, for a careful choice of the boundaries $B_1$ and $B_0$ this effect is minimized, and consequently it does not significantly affect the results. 

 \begin{figure}
 \includegraphics[width=1\linewidth]{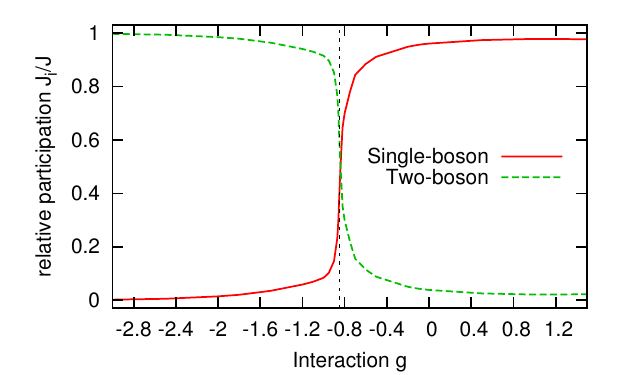}
 \caption{The ratios $J_1/J$ and $J_0/J$ as a function of $g$, showing the relative participation of single-particle tunneling and pair tunneling, respectively, in the overall tunneling dynamics of the two-boson system. Two distinct regimes can be identified, corresponding to the dominance of either single-boson or pair tunneling. The dashed line at $g = -0.85$ corresponds to the position of the peak in $\chi(g)$ (see Fig.~\ref{Fig5}). As can be seen, it matches very well the location of the transition between the two regimes. Interaction strength $g$ is given in units of $\sqrt{\hbar^3\Omega_0/m}$.}
 \label{Fig6} 
\end{figure}

The resulting individual fluxes $J_1(t)$ and $J_0(t)$ vary in time. However, within the time window in which ${\cal P}_2(t)$ decays exponentially, the ratios $J_1/J$ and $J_0/J$ (indicating participations of the sequential and pair tunneling in the overall decay) are essentially constant. In Fig.~\ref{Fig6} we plot these ratios as a function of interaction $g$. It can be seen that for $g < -0.85$ the tunneling process is nearly completely dominated by pair tunneling, while for $g > -0.85$ it is dominated by single-boson tunneling, with a smooth but rapid transition between the two regimes. This result confirms that the change in the behavior of the decay rate at $g \approx -0.85$ (see Fig.~\ref{Fig5}) is connected to a very rapid activation of sequential tunneling. 

In tunneling experiments determining the exact proportion of multi-particle decay can pose significant difficulties \cite{zurn2013}. However, our result shows that the behavior of $\gamma(g)$ can be used for indirect detection of the transition of the system between the different decay regimes. Since the decay rate $\gamma(g)$ can be obtained experimentally quite easily \cite{zurn2013}, therefore by measuring it for different interactions $g$ and calculating the susceptibility $\chi(g)$, it is in principle possible to gain insight into the form of the decay process.

At this point it is worth noting that in principle, the decay rates could also be found via time-independent methods. For example, one can exploit the WKB approximation (see for example \cite{zurn2012,rontani2013}). However, it is known that the approximation is oversimplified and may give inaccurate results \cite{gharashi2015}. Indeed, in the problem under study we found that calculations performed in the WKB approximation framework yield an underestimated decay rate for the non-interacting system. In this work in all cases the decay rates are obtained from exponential fits to the probabilities extracted from the numerically exact, time-dependent dynamics.

\section{Three bosons case}
\label{sec:3boson}

Now let us apply the above methods to analyze the decay of a three-boson system. The overall dynamics in this case still obeys the law of exponential decay. The survival probability $\mathcal{P}_S(t)$ is mimicked by the probability of finding all three bosons in the well, ${\cal P}_3(t)$. The probability ${\cal P}_3(t)$ is calculated by integrating the density $\rho(x_1,x_2,x_3,t)=|\Psi(x_1,x_2,x_3,t)|^2$ over the region $\mathbf{P}_3$ ($x_1,x_2,x_3 < x_w$) and the decay rate $\gamma$ is obtained from an exponential fit to ${\cal P}_3(t)$.

\begin{figure}
 \includegraphics[width=1\linewidth]{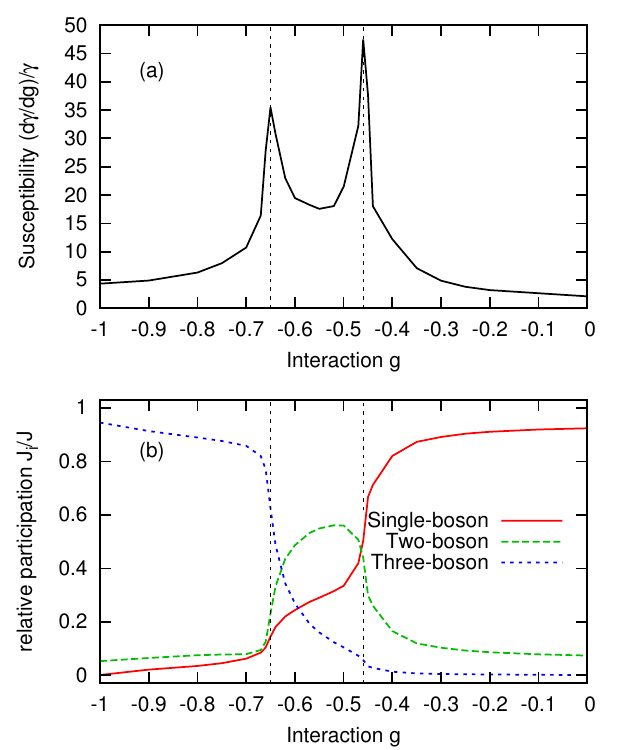}
 \caption{(a) The susceptibility $\chi(g) = \gamma^{-1} (\partial \gamma / \partial g)$, obtained for the decay rate $\gamma(g)$ of ${\cal P}_3(t)$ in the three-boson system. (b) The ratios $J_2/J$, $J_1/J$ and $J_0/J$ as a function of $g$, showing the relative participation of the three decay processes: single-particle tunneling, two-particle tunneling and three-particle tunneling, respectively. Three distinct regimes can be identified in the system, corresponding to dominance of single-, two- or three-particle tunneling. Vertical dashed lines correspond to locations of maxima in $\chi(g)$, which very well match the points of transitions between the regimes. Interaction strength $g$ and susceptibility $\chi$ are given in units of $\sqrt{\hbar^3\Omega_0/m}$ and $\sqrt{m/\hbar^3\Omega_0}$, respectively.}
 \label{Fig7}
\end{figure}

A significant difference is that there are now essentially three different decay channels possible. In addition to the single-boson tunneling and two-boson tunneling known from the two-particle case, the initial state can also decay through the emission of three bosons simultaneously. In Fig.~\ref{Fig7}a we show the susceptibility $\chi(g)$ of the decay rate as a function of $g$. Unlike the two-boson case, where a single peak was visible in the susceptibility, for a three-boson system two sharp peaks are clearly visible, at $g \approx -0.65$ and $g \approx -0.46$. By analogy to the two-boson case, it is natural to associate these peaks (and the corresponding rapid change in $\gamma(g)$) with changes in the dominant decay mechanisms. This indicates that there should be three distinct regimes discernible in the three-boson case, as opposed to the two regimes of the two-boson case.

To verify this, we apply our previous approach for analyzing the role of different tunneling channels. We define $J(t)$ as the total probability flux out of $\mathbf{P}_3$ at time $t$. Then we divide the boundary $\partial\mathbf{P}_3$ into three parts and we decompose $J(t)$ into $J_2(t)$, $J_1(t)$ and $J_0(t)$ corresponding to one-, two-, and three-boson tunneling, respectively. As previously, the ratios $J_i(t)/J(t)$ turn out to be almost time-independent, and they can be used to characterize the participation of the different tunneling mechanisms. In Fig.~\ref{Fig7}b we show the values of $J_i/J$ for different interactions $g$. It is clear that the locations of the peaks in $\chi(g)$ coincide with the transition points between three different regimes. For $g < -0.65$ the decay is dominated by three-boson tunneling. For intermediate interactions, $-0.65 < g < -0.46$, the two-particle tunneling is prominently present along with a nonvanishing contribution from three- and single-boson tunneling. Finally, for $g > -0.46$, single-boson tunneling constitutes the dominant decay channel. Notably, the second of these regimes is characterized by a non-negligible contribution of all three channels, rather than being completely dominated by a single decay process. This is an essential difference from the two-boson case, for which the two separate regimes are characterized by a total dominance of one particular channel. Note that this property cannot be captured by observation of the ${\cal P}_3(t)$ decay solely, while it is clearly visible when appropriate fluxes are considered. 

\section{Simple phenomenological treatment}
\label{sec:treatment}

\begin{figure}
 \includegraphics[width=1\linewidth]{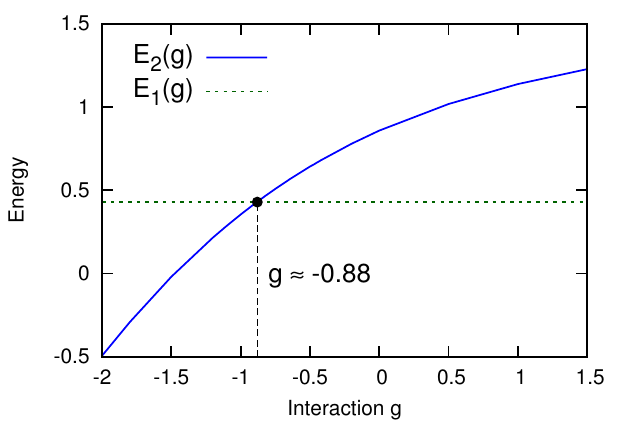}
 \caption{The initial energy $E_2(g)$ of the two-boson system (blue, solid), depending on interaction strength $g$, and the energy $E_1(g)$ of a single particle (green, dotted). The dashed vertical line indicates the interaction strength $g \approx -0.88$ above which $E_2(g) > E_1(g)$. Energy and interaction strength $g$ are given in units of $\hbar\Omega_0$ and $\sqrt{\hbar^3\Omega_0/m}$, respectively.}
 \label{Fig8}
\end{figure}

\begin{figure}                                                                              
 \includegraphics[width=1\linewidth]{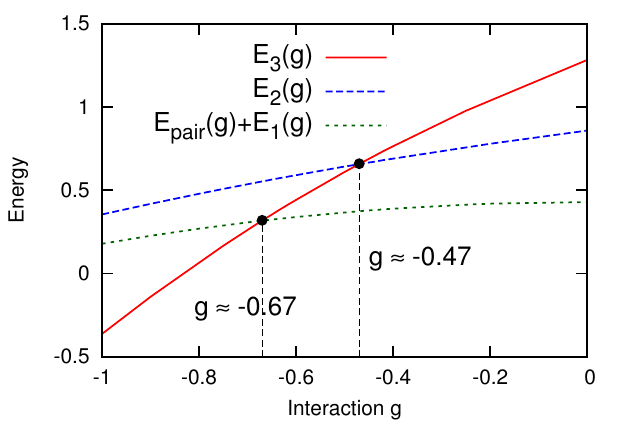}
 \caption{The initial energy $E_3(g)$ of the three-boson system (red, solid), the initial energy $E_2(g)$ of the two-boson system (blue, dashed), and the bound pair energy plus the energy of a single boson $E_{pair}(g) + E_1(g)$ (green, dotted), depending on interaction strength $g$. Vertical dashed lines indicate: the interaction strength $g \approx -0.47$ above which $E_3(g) > E_2(g)$, and the interaction strength $g \approx -0.67$ above which $E_3(g) > E_{pair}(g) + E_1(g)$. Energy and interaction strength $g$ are given in units of $\hbar\Omega_0$ and $\sqrt{\hbar^3\Omega_0/m}$, respectively.}
 \label{Fig9}
\end{figure}

In order to give an intuitive explanation of the results obtained, let us employ a simple theoretical description. At time $t = 0$, the $N$-boson trapped system has some energy $E_N(g)$. After one boson escapes the trap, the energy of the bosons remaining in the trap is equal to $E_{N-1}(g)$. Since the energy of the escaping boson cannot be negative (it is an almost free particle), the single-boson tunneling is possible only when $E_N(g) \ge E_{N-1}(g)$. Consequently, in the two-boson system under study ($E_1(g) \approx 0.43$ independently of $g$), we find that the single-boson tunneling condition $E_2(g) \ge E_1(g)$ is satisfied for $g \ge -0.88$ (see Fig.~\ref{Fig8}). This is remarkably close to the previously found transition point $g \approx -0.85$. In the case of three bosons, the corresponding condition $E_3(g) \ge E_2(g)$ is satisfied for $g \ge -0.47$ (see Fig.~\ref{Fig9}). The result is again very close to the previously found transition point $g \approx -0.46$, below which single-boson tunneling is suppressed. 

In the case of pair tunneling, the situation is more complicated, since one has to take into account the total energy of the tunneling pair. As noted in \cite{esebbag1993}, the internal energy of a freely moving bound pair is $E_{pair}(g) = -g^2/4$ (in natural units of the problem). Accordingly, pair tunneling is possible when $E_N(g) \ge E_{N-2}(g) - g^2/4$. Applying this result to the three-boson system, we find (as shown in Fig.~\ref{Fig9}) that the two-boson tunneling condition is satisfied for $g \ge -0.67$. This agrees very well with the transition point $g \approx -0.65$ below which pair tunneling is suppressed. Extending this phenomenological treatment to tunnelings of larger numbers of particles is straightforward. 

Note, however, that this simplified description is limited, since it cannot predict the relative importance of the different available tunneling channels. For example, within this approach one cannot predict that for intermediate interactions in the $N=3$ case the decay is governed by all three channels. However, the positions of transitions between different regimes can be predicted with very good precision.

\section{Role of the potential shape}
\label{sec:potential-shape}

A key feature of the studied potential is that it is constant in the region outside the well. This makes the phenomenological treatment described above possible, since precise energetic conditions for specific tunneling channels can be formulated. However, potentials used in experimental work may also take other forms. For example, in the Heidelberg experiment \cite{zurn2013} an external trapping potential ramps down outside the barrier and is not bounded from below at infinity ($x \rightarrow \infty$). Obviously in such a case our simplified phenomenological treatment breaks down, as there is no longer a specific lower bound for the energy of escaped particles. In consequence, a question arises how much the properties found previously are affected by the shape of the potential. To answer this question, we have analyzed the dynamics of two bosons in a modified potential, which is not bounded from below outside the well. We model this situation with the modified potential $V'(x)$ given by (\ref{eq:potential-unbounded}),

\begin{figure}
 \includegraphics[width=1\linewidth]{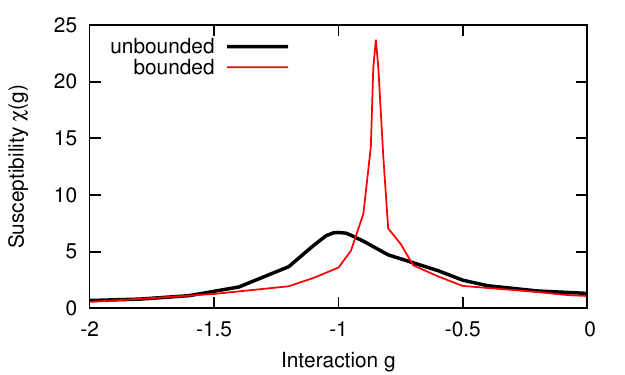}
 \caption{The susceptibility $\chi(g) = \gamma^{-1} (\partial \gamma / \partial g)$ for the modified, unbounded from below potential $V'(x)$ (thick black), compared to $\chi(g)$ in the bounded from below potential $V(x)$ (thin red). Due to the lack of precisely defined energy conversation conditions, the sharp peak becomes much more diffused in the unbounded potential. The maximum moves towards stronger attractions, since bosons with below-zero energies can now escape the well and the suppression of single-boson tunneling is diminished. Interaction strength $g$ is given in units of $\sqrt{\hbar^3\Omega_0/m}$, susceptibility in units of $\sqrt{m/\hbar^3\Omega_0}$.}
 \label{Fig10}
\end{figure}

On Fig. \ref{Fig10} we show the susceptibility $\chi(g)$ of the two-boson system obtained in the unbounded potential $V'(x)$, contrasted with $\chi(g)$ obtained for the bounded potential $V(x)$. It can be seen that the peak in susceptibility is still present, but it is significantly smaller and wider. This is an indication that the rapid, sharp transition observed earlier for $V(x)$ is significantly smoother for $V'(x)$. This follows naturally from the fact that strict energy conversation conditions can no longer hold when the potential energy beyond the barrier has no constant value. Additionally, since the potential takes on negative values (with respect to the local minimum of the well), bosons with negative energies can escape the well without violating the conversation of energy. This slows down the suppression of the single-boson tunneling for strongly attractive systems. Consequently, the maximum of $\chi(g)$ can be seen to move towards greater attractive interactions. 

The results indicate that the previously observed rapid transition between the different regimes is tied to the existence of precise energetic conditions for the different tunneling processes. In an unbounded potential, the separation into two distinct regimes is much less clear. 

\section{Conclusion}
\label{sec:conclusion}

We have analyzed the decay of a two- and three-boson system trapped in a one-dimensional potential well. In particular, we have investigated how the nature of the overall decay process changes as the interaction strength $g$ is changed. We show that the system undergoes transitions between several distinct regimes characterized by the dominance of different decay processes. Each such transition is reflected by a change in the behavior of the overall decay rate $\gamma(g)$ as $g$ is changed. This is seen clearly in the susceptibility $\chi(g) = \gamma^{-1} (\partial \gamma(g) / \partial g)$. 

For the two-boson system, we find a simple transition between two regimes, one dominated by single-boson tunneling and one by pair tunneling. For the three-boson system, we find three distinct regimes, characterized by the dominance of one-, two- and three-boson tunneling channels. Importantly, we show that the intermediate regime with two-boson tunneling has non-negligible contributions from the other tunneling processes. Thus, one cannot neglect these processes when studying the dynamics in this regime. This result cannot be obtained basing on the rate $\gamma(g)$ alone, but it can be found by analyzing the appropriate fluxes of probability between various regions of configuration space. The interaction strengths for which the transitions occur can be approximately determined via a simple theoretical description. While we only present results for two and three bosons, the overall approach can be quite easily extended to a larger number of particles. 

We have also investigated the effects of the asymptotic form of the external potential at infinity. We find that the transition between the two regimes is sharp and clearly discernible when the potential remains bounded at infinity ($x \rightarrow \infty$). In contrast, when the potential is not bounded from below, the transition is much smoother and the individual regimes are less distinct.

This work builds upon existing research on escaping few-body systems. In particular, the distinction between regimes dominated by single- and pair-tunneling was considered already \cite{zurn2013,rontani2013}. Moreover, in \cite{gharashi2015} the analysis of flux dynamics was used to estimate the relative importance of single- and two-particle tunneling. However, we expand upon these results by applying the method to a three-particle system, showing that flux analysis can be applied to systems with a greater number of particles. Furthermore, we provide a detailed description of how the separate tunneling channels contribute to the overall decay process. In this way we show that in an in-depth analysis of bosonic tunneling dynamics, several distinct decay channels must be jointly taken into account to obtain accurate results. We also show that the transitions between the individual regimes can be indirectly detected through measurement of just one observable: the probability of finding all bosons in the trap.

The results are potentially relevant to experimental practice. Determining the decay rate $\gamma(g)$ for different interactions is possible experimentally, and the peaks in its susceptibility $\chi(g)$ can be used to indirectly detect a transition between specific tunneling regimes, providing an alternative method of probing the system dynamics. 

\section{Acknowledgments}

The authors are grateful to Andrea Bergschneider and Przemys{\l}aw Ko{\'s}cik for their fruitful suggestions and questions. This work was supported by the (Polish) National Science Center Grant No. 2016/22/E/ST2/00555.

\end{document}